\documentclass[floatfix,%
 reprint,
 amsmath,amssymb,
 aps,prl,
 showkeys,
]{revtex4-2}
\usepackage[utf8]{inputenc}
\usepackage{graphicx}%
\usepackage{caption}
\usepackage{subcaption}
\usepackage{siunitx}
\usepackage{adjustbox}
\usepackage{xr}
\usepackage{dcolumn}%
\usepackage{bm}%
\usepackage{hyperref}%
\usepackage[mathlines]{lineno}%
\usepackage{float} %
\usepackage[labelsep=period]{caption}
\usepackage{appendix}

\usepackage[dvipsnames]{xcolor}
\usepackage{caption}
\definecolor{mgreen}{rgb}{0.1,0.7,0.1}

\begin{document}

\title{Memories of amplitude and direction coexist and compete in non-Brownian suspensions}%
\thanks{E-mail: ssp5361@psu.edu}%
\author{
Surendra Padamata$^{1}$ Nathan C.\ Keim$^{1}$
\\
$^{1}$Department of Physics, The Pennsylvania State University, University Park, PA 16802, USA
}

\date{\today}
\begin{abstract}
       Steadily shearing a non-Brownian suspension forms a memory of direction, while shearing back and forth forms a memory of amplitude. Each memory is evident in the system's response to further shear, exemplifying its strong history-dependence. By combining the steady and oscillatory experiments, we show these memories are distinct but intersecting aspects of the same non-equilibrium physics: they can coexist, yet a specific amplitude suppresses directional memory and makes the system symmetric. Combined with prior results from disordered solids, our work presents a simple motif for limited memory capacity in non-equilibrium matter.
\end{abstract}


\maketitle

Physical systems can store information about their thermal or mechanical histories. These memories take many forms and can be surprisingly detailed. A full accounting of what information is retained and lost can be a fingerprint of the specific non-equilibrium nature of a ferromagnet, a piece of sandstone or glass, or a charge density wave conductor~\cite{RevModPhys.91.035002,PaulsenKeimSolidsARCMP2023,Dasent2025Faultzone}. In these studies, one behavior is so simple and ubiquitous that in some contexts it seems trivial: the memory of the most recent direction of driving \cite{10.1122/1.549944, PhysRevLett.93.088001, PhysRevE.82.026104, Kamani2025Memory}.%

\begin{figure}[b!]
    \centering
    \begin{subfigure}[]{0.15\textwidth}
        \centering
        \includegraphics[width=\textwidth, height=1.6\textwidth]{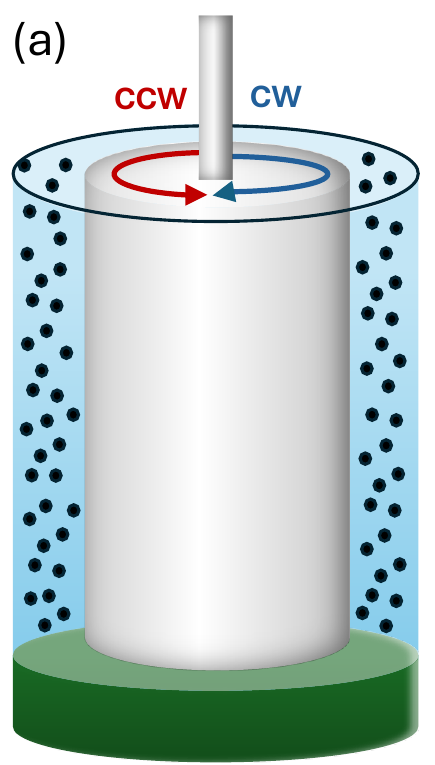}
        \label{fig:a}
    \end{subfigure}
    \begin{subfigure}[]{0.28\textwidth}
        \centering
        \begin{subfigure}[b]{\textwidth}
            \centering
            \includegraphics[width=\textwidth]{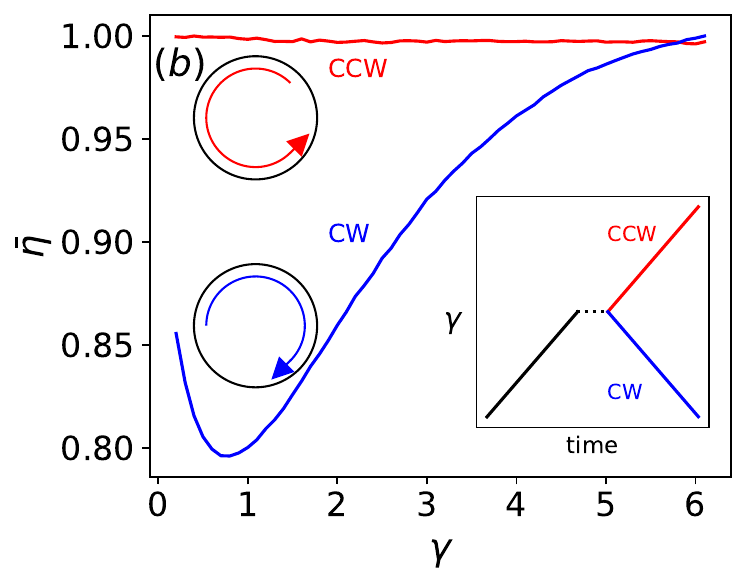}
        \end{subfigure}
        \begin{subfigure}[]{\textwidth}
            \centering
            \includegraphics[width=1.1\textwidth]{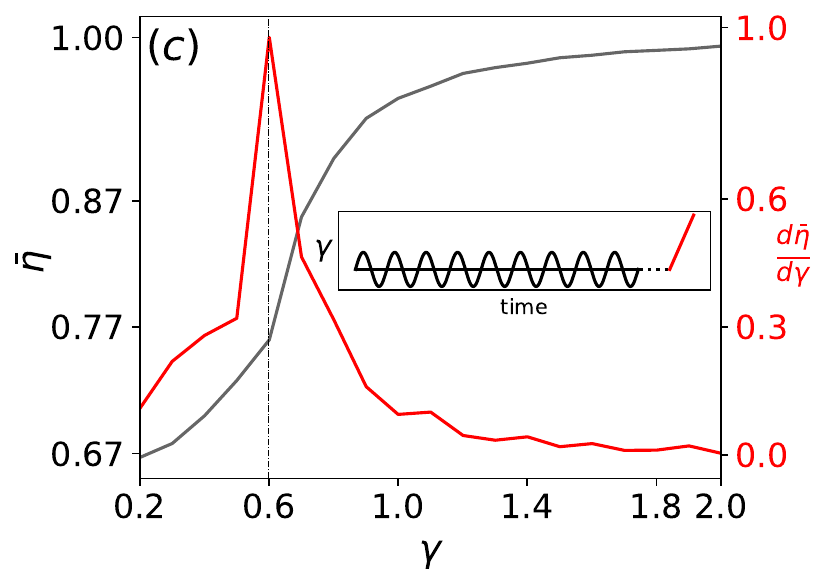}
        \end{subfigure}
    \end{subfigure}
    \caption{Writing and reading memories in non-Brownian suspensions.
    \textbf{(a)} Experimental setup. Circular arrows are shear directions.
    \textbf{(b)} Memory of direction. \emph{Inset}: Protocol. Initial steady shear (black) in CCW direction precedes readout (red or blue).
    \emph{Main panel}: Normalized viscosity (see text) during readouts. %
    Resuming same direction (red) yields same viscosity; reversing (blue) causes a drop.
    \textbf{(c)} Memory of amplitude. \emph{Inset}: Oscillatory protocol with amplitude $\gamma_T = 0.6$ and readout (red).
    \emph{Main panel}: Viscosity during readout (black). Derivative (red curve) peaks at $\gamma_T$.}
    \label{fig:full}
\end{figure}

One well-known system with directional memory is a suspension of particles in viscous liquid, as in the simple experiment in Fig.~\ref{fig:full}(a, b). The spherical particles occupy about 30\% of the total volume, and they are non-Brownian and neutrally buoyant. The particles and their interactions make the suspension more viscous than the liquid alone. In Fig.~\ref{fig:full}(b), a sample is sheared steadily in one direction and brought to rest. If shear resumes in the same direction, the viscosity is unchanged. However, when the shear is reversed, the overall viscosity drops by about 20\%---meaning that most of the contribution from interactions among particles is suppressed~\cite{1972JFM....56..401B}. The viscosity then gradually returns to its steady-state value. Evidently, shearing made the suspension's structure anisotropic, so that the short-range interactions between particles are much weaker when shear is reversed~\cite{GadalaMaria1980ShearInduced, 10.1122/1.549944}. Observing a new viscosity when revisiting a previous strain defies the expectation that when inertia is negligible, flow should be kinematically reversible. This memory is a glimpse of the deep challenges of understanding suspension structure during flow~\cite{Guazzelli2012Suspension}.
However, this example presents a more immediate puzzle: is it possible to erase the memory of one direction without forming the memory of another? When sedimentation and Brownian motion are negligible, must suspensions always carry memories of flow?

The complex interplay of structure and flow also leads to a second type of memory. Over many cycles of oscillatory shear, the suspension forms memories of strain amplitudes. Figure~\ref{fig:full}(c) shows that a stored amplitude $\gamma_T$ increases the viscosity sharply as the suspension is sheared past that strain~\cite{PhysRevLett.107.010603, PhysRevLett.113.068301}. %
The suspension attains kinematic reversibility, with particles tracing the same path on both the forward and reverse cycles. Viewed stroboscopically, their positions stop changing unless the amplitude exceeds $\gamma_T$~\cite{Pine2005-yg,PhysRevLett.113.068301}. A similar picture for amplitude memory arises in systems like charge density wave conductors, amorphous solids, and vortices in superconductors~\cite{RevModPhys.91.035002}. There is a critical amplitude $\gamma_c$ at which the number of cycles needed to reach reversibility diverges. Past this transition, a reversible state is unreachable, and memories are suppressed.

In this Letter, 
we consider directional and amplitude memory together. 
We are inspired by research on cognitive working memory~\cite{pnas.1315171111, Hachen2021-pi}, where interactions between different kinds of memory, whether facilitative or inhibitive~\cite{McClelland1995,Dudai2004}, can support function.
In suspensions, we observe that the two memories coexist and influence how each other is read out, but they become mutually exclusive when strains approach $\gamma_c$---implying that the structural motifs that encode these memories are distinct yet overlapping.
Strikingly, memory of direction, memory of amplitude, and a reversible-to-irreversible transition are also found together in amorphous solids under shear, despite very different microscopic physics, suggesting a common way that uniaxial %
driving navigates configuration space in %
non-equilibrium systems. Our work establishes a more unified picture of history-dependence in non-inertial suspensions, and provides a non-trivial example of finite memory capacity in nonliving and living systems.

\emph{Experiments ---} We create a neutrally buoyant suspension with a 30\% volume fraction of PMMA particles (Cospheric, LLC) with a roughly Gaussian distribution of diameters around 114.7~$\mu$m, with $>98\%$ between 106 and 125~$\mu$m. The liquid is a solution of Triton X-100 (75.08\%), water (12.30\%), and zinc chloride (12.62\%)%
~\cite{Corte2008-op, PhysRevLett.113.068301,Krishnan_Beimfohr_Leighton_1996}. A noticeable density gradient appears after 1--2 months, with a density mismatch of $\sim$$10^{-4}$ g/cm$^3$. Sedimentation is therefore negligible during each $\sim$1-hour test, and stored memories persist for at least a day. We shear the suspension between concentric cylinders with radii 14 and 22~mm %
in a rheometer (TA Instruments DHR-20). Shear strains %
and strain rates are calculated at radius $\sim$18~mm. The outer cylinder's temperature is held at 22~$^\circ$C. To minimize boundary effects, the suspension floats on low-viscosity fluorinated oil (Fluorinert FC-70, 3M). The typical strain rate is $\dot{\gamma} = 0.1$ s$^{-1}$. In this regime, we neglect inertia (Reynolds number $\sim 10^{-3}$) and Brownian motion (Peclet number $\sim 10^9$). Lowering strain rates by a factor of 10 does not change our results~\cite{supplemental}.

\begin{figure}[]
    \centering
    \subfloat{
    \includegraphics[width=0.22\textwidth]{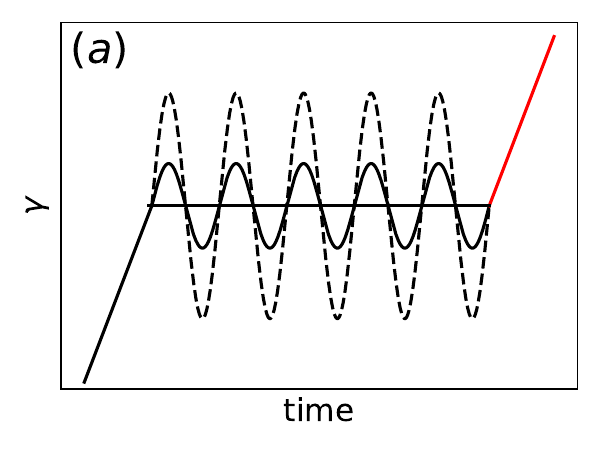}
    }
    \subfloat{
    \includegraphics[width=0.22\textwidth]{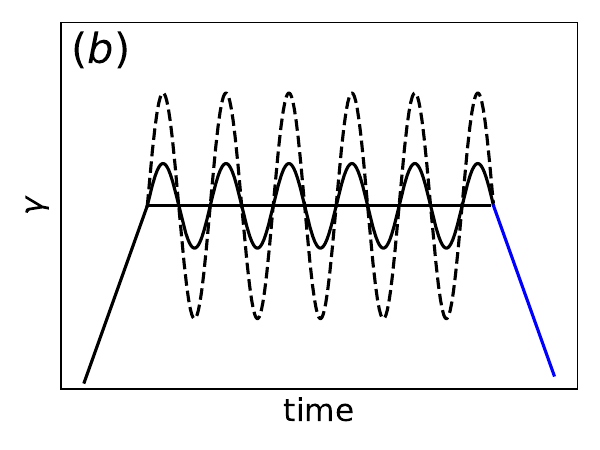}
    }
     \hfill
    \subfloat{
    \includegraphics[width=0.45\textwidth]{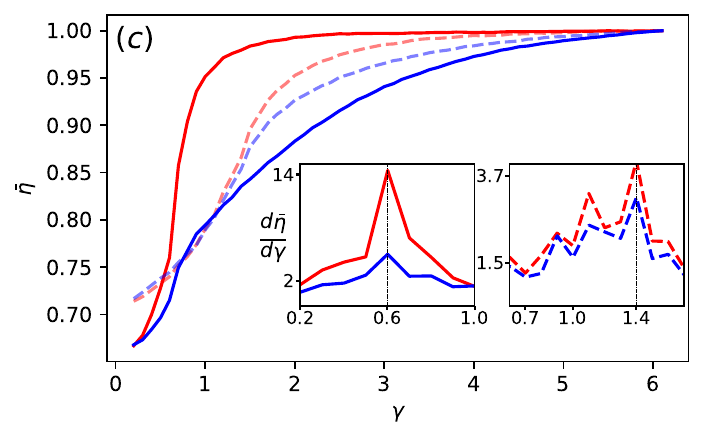}
    }
    \caption{Memories of direction and amplitude can coexist. 
    \textbf{(a)} Schematic of protocols that combine memories. Readout (red) matches direction of preparation. Oscillatory portion shows amplitudes $\gamma_T=0.6$ (solid) and $\gamma_T=1.4$ (dashed). Only five cycles are shown for clarity (see text).
    \textbf{(b)} Protocol with readout (blue) in opposite direction. To avoid introducing a reversal at $\gamma = 0$ we add a half-cycle.
    \textbf{(c)} Normalized viscosity during readouts. Red and blue curves are from protocols (a) and (b) respectively; solid and dashed lines represent the two $\gamma_T$. \emph{Insets}: Derivative of viscosity shows peaks at $\gamma_T$.
    }
    \label{fig:modap_vis_dvis}
\end{figure}

Figure~\ref{fig:modap_vis_dvis}(a, b) show experimental protocols that combine direction and amplitude memories, for training amplitudes $\gamma_T = 0.6$ and $1.4$. We prepare the system with steady shear at 0.1~s$^{-1}$ for 200--400~s to erase prior memories. For simplicity, here we present protocols with counterclockwise preparation ($\dot{\gamma} > 0$); inverting our protocols ($\gamma \to -\gamma$) gives equivalent results~\cite{supplemental}. 
For $\gamma_T = 0.6$ and $1.4$, we apply oscillatory training with a period of 25~s for 10 cycles, to reach a steady state~\cite{supplemental,Corte2008-op}. Finally, we read out memories with steady shear at 0.1~s$^{-1}$ in either direction. We use ``match'' protocols in which the direction at the end of training matches the direction of readout, which we achieve by adding an extra half-cycle of training as needed. This method avoids creating a turning point at $\gamma = 0$. During readout, we sample viscosity at intervals of 0.1 in strain. The training ends at a small nonzero strain so that readout then passes through $\gamma = 0$. This offset is $|\gamma| = 0.11 \gamma_T$ for the Fig.~\ref{fig:modap_vis_dvis}(a) protocol and $|\gamma| = 0.071 \gamma_T$ for Fig.~\ref{fig:modap_vis_dvis}(b). We report normalized viscosity $\bar \eta$ by dividing the raw viscosity by its value at the end of readout ($\gamma \approx 6$), typically  13--15~Pa~s, to compensate for an apparent slow variation in liquid viscosity, possibly due to ambient temperature changes~\cite{supplemental}. The derivative for each $\gamma_i$ is computed as $(\bar \eta_{i+1}-\bar \eta_{i}) / ({\gamma_{i+1}-\gamma_{i}})$, meaning that the rise in $\bar \eta$ at $\gamma > \gamma_T$ will be assigned to $\gamma_T$.

\emph{Results ---} Figure \ref{fig:modap_vis_dvis}(c) plots the viscosity and its derivative during readout for each amplitude and direction in Fig.~\ref{fig:modap_vis_dvis}(a, b). The memory of direction survives oscillatory shear: as in Fig.~\ref{fig:full}, viscosity is lower when the initial steady shear and readout are opposite (blue curves). The insets show that the memory of amplitude is present in both directions, but strongest in the direction of initial preparation. Each readout thus contains information about both types of memory. 
We note that due to the small strain offsets at the beginning of readouts, the sampling resolution (Fig.~S6 in \cite{supplemental}), and the possible perturbative effects of the memory of direction, we cannot say whether the peak in $d \bar \eta / d \gamma$ indicates $\gamma_T$ exactly, but there is a close correspondence. When we independently vary the positive and negative turning points of strain in Appendix~A, opposite readouts reveal two strain memories, as proposed earlier~\cite{PhysRevLett.113.068301, PhysRevE.88.032306}. Thus while our protocols were chosen for simplicity, a more general approach might consider the history of arbitrary turning points~\cite{RevModPhys.91.035002}.

\begin{figure}[]
    \centering
    \subfloat{
    \includegraphics[width=0.45\textwidth]{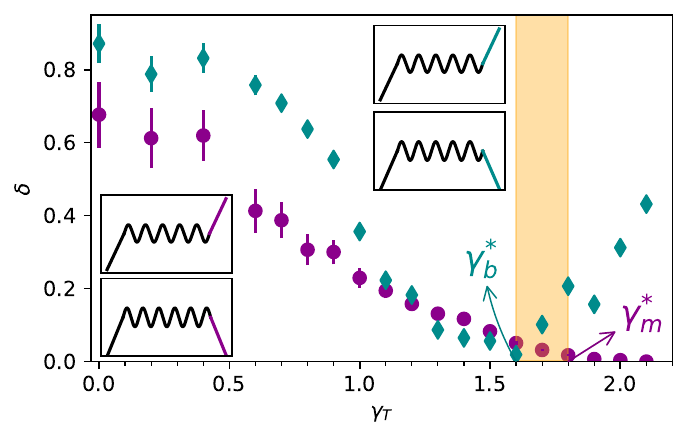}
    }
    \caption{Effect of oscillatory shear on memory of direction. 
    \emph{Magenta circles:} Asymmetry $\delta$ corresponds to area between blue and red curves in Fig.~\ref{fig:modap_vis_dvis}c (protocols reproduced in lower-left insets). With sufficiently large amplitude $\gamma_T$, $\delta = 0$ and there is no trace of the direction of initial shear.
    \emph{Blue diamonds:} Results when oscillatory shear always ends in the same direction as preparation (protocols in upper-right insets). Amplitude $\gamma_T \approx 1.6$ erases memory of direction; $\gamma_T > 1.6$ forms a new one.
    }
    \label{fig:asymmetry_measure_norm}
\end{figure}

The viscosity curves in Fig.~\ref{fig:modap_vis_dvis} also suggest a reciprocal effect: increasing the oscillatory shear amplitude $\gamma_T$ makes the CW and CCW readout curves more similar (see Fig.~S4 for more examples~\cite{supplemental}). To explore this idea, we define $\delta$ to be the integrated area between the CW and CCW curves, yielding a measure of asymmetry. 
After removing the transient from the start of each readout, we average $\bar \eta$ from 3 trials for each protocol, and 3 more trials from the equivalent inverse protocol~\cite{supplemental}. After computing this aggregate $\delta$, we estimate random uncertainty by computing $\delta$ for the 36 possible pairings of individual trials and taking the standard deviation. Prior experiments~\cite{Corte2008-op} found that at larger $\gamma_T$, more cycles are needed to reach a kinematically reversible steady state, so for $\gamma_T \geq 1.4$ we apply up to 40 cycles, even though the additional cycles do not affect $\delta$~\cite{supplemental}. %

The magenta circles in Fig.~\ref{fig:asymmetry_measure_norm} show $\delta$ from the ``match'' protocols of Fig.~\ref{fig:modap_vis_dvis} at a range of training amplitudes $\gamma_T$.
We find that $\delta$ falls with increasing $\gamma_T$, and is nearly zero for $\gamma_T \ge 1.8$, which we refer to as $\gamma^*_m$. The protocols are mirror inverses of each other ($\gamma \to -\gamma$), except for the first steady shear---meaning that $\delta$ measures only the influence of that initial preparation. This memory is lost for $\gamma_T \ge \gamma^*_m$. 

Figure~\ref{fig:asymmetry_measure_norm} plots a second set of experiments as blue diamonds.  Instead of matching the direction at the end of oscillatory training with the direction of readout (as in Fig.~\ref{fig:modap_vis_dvis}), we introduce ``break'' protocols shown in the right-hand insets of Fig.~\ref{fig:asymmetry_measure_norm}: Oscillatory training always ends in the same direction, and only the direction of readout differs. In this case, $\delta$ measures asymmetry at the end of training, regardless of its origin. This $\delta$ has a minimum at $\gamma_T \approx 1.6$, which we refer to as $\gamma^*_b$. The subsequent rise corresponds to a new memory of the final direction of training, consistent with the large-amplitude limit in which oscillatory training is effectively steady shear in alternating directions. 

Remarkably, the ``break'' data suggest that there exists an amplitude $\gamma_b^*$ that is both large enough to erase the direction of initial shear, and too small to impose a new direction. At this amplitude the trained suspension behaves symmetrically. The different $\gamma_T$ at which the two curves reach zero ($\gamma_m^*$, $\gamma_b^*$) and the different $\delta$ values at $\gamma_T = 0$ may be due to the ``break'' experiments having a higher effective volume fraction, consistent with the slow sedimentation we observe~\cite{PhysRevFluids.3.124303, supplemental}.

\begin{figure}[]
    \centering
    \subfloat{
    \includegraphics[width=0.45\textwidth]{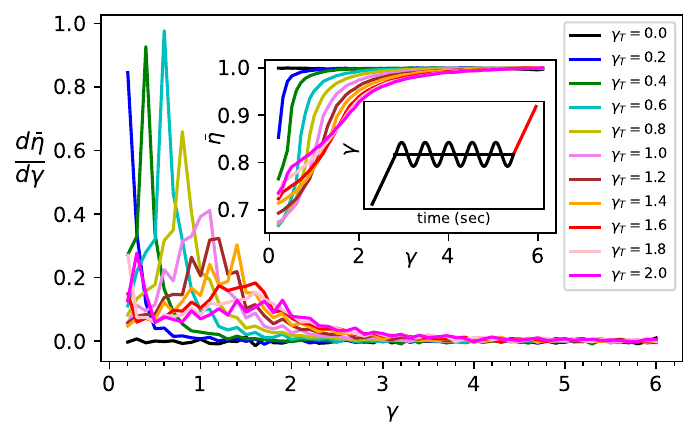}
    }
    \caption{
    Memory peaks for training amplitudes from 0.2 to 2. Beyond 1.6, the amplitude memory is not discernible from the background even after 40 cycles of training. Black curve with $\gamma_T = 0.0$ (no training) is equivalent to resumption of steady shear in Fig.~\ref{fig:full}(b). Insets: Viscosity curves and abbreviated protocol. 
    }
    \label{fig:all_amplitudes}
\end{figure}

As the directional memory of the suspension grows stronger for $\gamma_T > \gamma^*_b$, Fig.~\ref{fig:all_amplitudes} shows that the signatures of amplitude memories become weaker and broader. We take $\gamma^*_a \approx 1.6$ to be the largest $\gamma_T$ that forms a discernible memory; for higher $\gamma_T$, the peak location is indistinct and does not match $\gamma_T$. Peaks are highest for $\gamma_T = 0.6$---in contrast with amorphous solids, where memories are strongest near the largest amplitude that can be remembered~\cite{PhysRevLett.122.158001}.  

Lastly, we return to the case of no training in Fig.~\ref{fig:full}(b). The viscosity after strain reversal has an extended region with large, near-constant slope, beginning with an inflection point at $\gamma_{d} \approx 1.8$. While this corresponds to a peak in $d \bar\eta / d\gamma$ (Appendix B), the peak is associated not with any turning point at that strain, but with the memory of direction and its turning point at $\gamma = 0$. Thus $\gamma_d$ is a characteristic strain scale for the onset of particle interactions after reversal.

\emph{Discussion ---} Non-Brownian suspensions can store directional and amplitude memories, as previously demonstrated in isolation. Our experiments establish that these two kinds of memory compete and can even exclude one another: the suspension always carries information about its flow history, but the amount and type vary. 

We can begin to understand this competition by considering interparticle interactions. First, we observed that memories of amplitude are weaker when the directions of preparation and readout are opposite (Fig.~\ref{fig:modap_vis_dvis}). Memory of direction implies weaker interactions upon reversing shear, so that the flow is temporarily closer to the dilute limit of kinematic reversibility. Because these interactions are suppressed \emph{partially}, new memories can be formed even at small amplitudes and against the direction of preparation---but these memories must be weaker than those formed in the same direction as preparation.
Second, we observed that a memory of amplitude weakens the memory of direction. Since forming any memory requires irreversibly remodeling the suspension structure, we surmise that oscillatory training disrupts the strongly anisotropic structure that was built up by steady shear---but at small amplitude this disruption is likewise partial.

As the strain of oscillatory or steady shear increases, the competition intensifies and memories can be lost completely. Remarkably, our results point to a common strain scale, $\sim$1.6 in our sample, which is evident as:
(i) $\gamma^*_m$, the minimum strain amplitude for erasing the direction of earlier shearing;
(ii) $\gamma^*_b$, the minimum amplitude to form a new memory of direction; 
(iii) $\gamma^*_a$, beyond which amplitude memories are indistinct; 
(iv) $\gamma_d$, the location of the inflection point in viscosity after reversing steady shear;
(v) $\gamma_c$, the critical amplitude past which the system cannot become kinematically reversible, which Cort\'e et al.\ found to be $1.6$ at this volume fraction of 30\%~\cite{Corte2008-op}. All of these similar strain scales are measured relative to turning points of shear, and all are non-equilibrium behaviors that require irreversible interactions between particles.

Hints about the origin of this strain scale come from the role of particle volume fraction $\phi$, which earlier works varied for two of the phenomena: Parsi and Gadala-Maria~\cite{10.1122/1.549944} and Refs.~\citenum{PhysRevLett.107.208302, 10.1122/1.4766597,2023AcRhe..62..253L} studied the strain over which viscosity recovers after reversing steady shear, %
which we characterize in this work with $\gamma_d$; while Pine et al.~\cite{Pine2005-yg} measured $\gamma_c$ for oscillatory shear. Both strains scaled as $\phi^{\alpha}$, where $\alpha = -2.0 \pm 0.2$. A more recent study of steady and oscillatory shear also showed strain-dependent viscosity or loss modulus that were consistent with this scaling~\cite{LIN2015228}.
While our work is at a single $\phi$, and the scaling of our $\gamma^*$ remains to be tested, these prior results offer clues as to the mechanism for the present behavior.
The prevalence of neighbors around a particle scales as $\phi$, and so, all else being equal, the strain required to bring one of those neighbors into close proximity should scale as $\phi^{-1}$. The observed scaling $\sim \phi^{-2}$ suggests that the strain scale instead comes from higher-order crowding effects: limits on how anisotropic or heterogeneous a particle's environment can become, which in turn limit the scale at which hydrodynamic interactions and viscosity can be enhanced or suppressed. This point is underscored by a successful many-body model that attributes $\gamma_c$ to a percolation transition~\cite{Corte2008-op, PhysRevE.79.061108}. More directly, we note that a theoretical ensemble of two-particle systems can easily hold both direction and amplitude memories at \emph{any} strain~\cite{PhysRevE.88.032306}, suggesting that many-body interactions are responsible for the competition between memories in our experiments at large strain. %

We also note that the origin of irreversibility is not firmly established, but that it likely involves interactions between microscopic features on particles' rough surfaces \cite{PhysRevE.75.066309}. Therefore, more %
mechanistic insights are likely to come from future experiments that measure the pair-correlation function $g(\vec r)$ in the plane of shear and observe its evolution, while systematically varying particle volume fraction, stiffness, and roughness---building on experiments with steady shear \cite{2023AcRhe..62..253L, 10.1122/1.549944, GadalaMaria1980ShearInduced}. 
Such studies may also deepen understanding of the reported hyperuniform structure at $\gamma_c$~\cite{PhysRevLett.115.108301, PhysRevLett.125.148001}, since hyperuniformity alone does not exclude a global anisotropy that encodes direction, or nearest-neighbor pairs that encode strain amplitude.

Recent studies of a 2D jammed, amorphous solid suggest a similar interplay between memories of direction and amplitude~\cite{Galloway2022-ok, PhysRevX.15.011043, KeimMedinaSciAdv2022}. Although particle interactions are very different (jammed and conservative, instead of dilute and dissipative), after oscillatory shear at amplitude $\lesssim 0.1$ the system generally has memories of both amplitude and the direction of preparation~\cite{RevModPhys.91.035002, PaulsenKeimSolidsARCMP2023}. As with suspensions, the direction is erased at larger amplitudes, where there is an apparent critical transition in which the system can no longer find a reversible steady state~\cite{PhysRevResearch.5.021001}. 
Studies with quenched elastoplastic models further show that amplitude memory depends on direction~\cite{kumar2024, mungan2025}. Memories of amplitude and direction, and critical reversible-irreversible transitions under cyclic driving, have also been observed separately in frictional granular solids and many more systems~\cite{toiya2004,benson2021,PaulsenKeimSolidsARCMP2023, PhysRevResearch.5.021001,Dasent2025Faultzone}.

Our present results in suspensions lead us to propose that the triad of amplitude memory, directional memory, and reversibility, organized around a critical point, is a generic motif worth further consideration. Our method is general and could reveal a unified picture in many more systems. 
These common behaviors are more fascinating because of the distinct physics of each system---the details of how amplitude and directional memories coexist and compete within the same reversible microstructure.

We thank Omri Barak, Justin Burton, Paul Lammert, Mathew Diamond, Muhittin Mungan, David Pine, Joseph Paulsen, and Michael Rubinstein for helpful discussions, and Sebanti Chattopadhyay and Julia Rice for technical assistance. This work was supported by the Human Frontier Science Program (Ref.\ No.\ RGP0017/2021).

\bibliography{bibliography}

\clearpage

\appendix

\section*{End Matter}

\section{Appendix A: Turning points}\label{appendix:Turning-points}

\begin{figure}[H]
    \centering
    \subfloat{
    \includegraphics[width=0.45\textwidth]{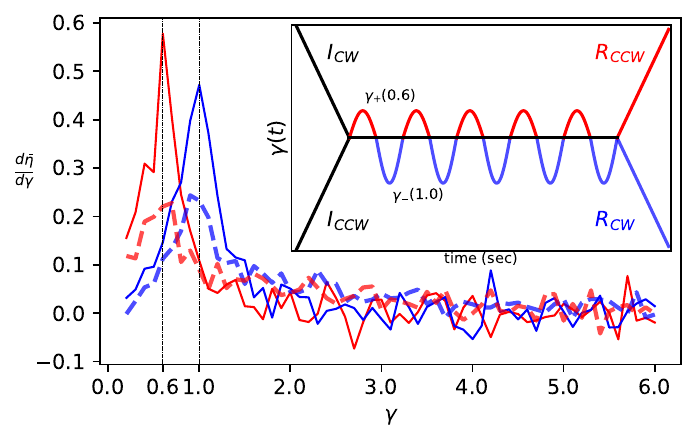}
    }
    \caption{\textbf{Turning points are remembered separately.} Plot shows first derivative of viscosity during readout. The red curve's peak corresponds to the memory of $\gamma_{+}$, while the blue curve's peak represents the memory of $\gamma_{-}$. The dashed red and blue peaks show $\gamma_{+}$ and $\gamma_{-}$ when the direction of the initial steady shear is reversed relative to their respective readout directions, which attenuates the memory of amplitude.}
    \label{fig:full_response}
\end{figure}

Figure \ref{fig:full_response} shows the first derivative of viscosities during readout. When initial preparation $I$ and the final readout $R$ are in the same counterclockwise direction ($I_{CCW}$ and $R_{CCW}$), we observe the positive turning point (red peak). When they are in the same directions but clockwise ($I_{CW}$ and $R_{CW}$), we observe the negative turning point (blue peak). When $I_{CW}$ and $R_{CCW}$ are opposite, we observe the positive turning point with an additional effect from the direction (dashed red peak). Similarly, for $I_{CCW}$ and $R_{CW}$, we observe the negative turning point with an additional effect from the direction (dashed blue peak). This implies four possible readouts of a given oscillatory protocol.

\section{Appendix B: Strength of amplitude memory}\label{appendix:Strength of amplitude memory}

\begin{figure}[H]
    \centering
    \subfloat{
    \includegraphics[width=0.45\textwidth]{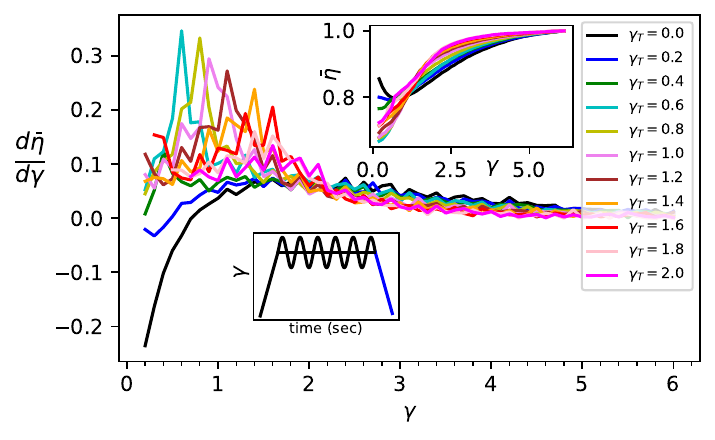}
    }
    \caption{\textbf{Memory peaks for training amplitudes 0--2 with preparation and readout in opposite directions}. The lower inset illustrates the protocol. The main plot demonstrates that beyond $\gamma_T = 1.6$, the amplitude memory becomes indistinguishable from the background even after 40 cycles of training. Each peak is smaller than its counterpart in Fig.~4 of the main text, where initial preparation and readout are in the same direction. As in that case, however, the strongest peak is for $\gamma_T = 0.6$. The black curve with $\gamma_T = 0.0$ (i.e.\ no oscillatory training) is equivalent to the derivative of the viscosity after reversing steady shear, plotted in Fig.~1(b) of the main text. Upper inset: normalized viscosity.}
    \label{fig:all_amplitudes_foreaft}
\end{figure}

Figure~\ref{fig:all_amplitudes_foreaft} is the counterpart to Fig.~4 in the main text, with readout in the opposite direction as initial steady shear. As expected, memory peaks are smaller due to the memory of direction, and the directional effect is strongest at small $\gamma_T$, and disappears near $\gamma_T \approx 1.6$. In both sets of experiments, the largest memory peak is for $\gamma_T = 0.6$. The black $\gamma_T = 0.0$ curve, corresponding to no oscillatory training, has a broad maximum that begins at $\gamma_d \approx 1.8$, consistent with the steady-shear reversal result in Fig.~1(b) of the main text. There has been a recent interest in quantifying the height of memory peaks with respect to $\gamma_T$, particularly in amorphous solids, as briefly mentioned in the main text \cite{PhysRevLett.122.158001}. In contrast to that work, we observe that the peak height reaches its maximum at a $\gamma_T = 0.6$, which is neither the minimum nor the maximum strain where amplitude memory is possible. Initially, the peak heights increase until $\gamma_T = 0.6$, reach a maximum, and then gradually decrease, eventually falling below the noise floor of the experiment. Interestingly, the directional memory does not alter this qualitative trend in the peaks.

\end{document}